# Seasonal variation of the mesospheric inversion layer, thunderstorms and mesospheric ozone over India

S. Fadnavis[1], Devendraa Siingh[1,2], G. Beig[1] and R. P. Singh[3]

[1]Indian Institute of Tropical Meteorology, Pune-411 008, India

[2]Current affiliation: University of Tartu, Institute of Environmental Physics, 18 Ulikooli Street, Tartu-50090, Estonia

[3]Vice-Chancellor, Veer Kunwar Singh University, ARA (Bihar)-80230, India

## Abstract

Temperature and ozone volume mixing ratio profiles obtained from the Halogen Occultation Experiment (HALOE) aboard the Upper Atmospheric Research Satellite (UARS) over India and over the open ocean to the south during the period 1991-2001 are analyzed to study the characteristic features of the Mesospheric Inversion Layer (MIL) at 70 to 85 km altitude and its relation with the ozone mixing ratio at this altitude. We have also analyzed both the number of lightning flashes measured by the Optical Transient Detector (OTD) onboard the MicroLab-1 satellite for the period April 1995 - March 2000 and ground-based thunderstorm data collected from 78 widespread Indian observatories for the same period to show that the MIL amplitude and thunderstorm activity are correlated. All the data sets examined exhibit a semiannual variation. The seasonal variation of MIL amplitude and the frequency of occurrence of the temperature inversion indicate a fairly good correlation with the seasonal variation of thunderstorms and the average ozone volume mixing ratio across the inversion layer. The observed correlation between local thunderstorm activity, MIL amplitude and mesospheric ozone volume mixing ratio are explained by the generation, upward propagation and mesospheric absorption of gravity waves produced by thunderstorms.

## 1. Introduction

One of the spectacular transient phenomena observed on certain days at mesospheric altitudes is the Mesospheric Inversion Layer (MIL) or thermal inversion layer at altitudes between 70 to 85 km. This was first reported by Schmidlin (1976) and later on by various workers into the measured temperature at low and mid latitudes by different techniques such as a Rayleigh lidar (Hauchecorne et al., 1987; Jenkins et al., 1987; Meriwether et al., 1998; Thomas et al., 2001; Kumar et al., 2001), a sodium lidar (She et al., 1995; States and Gardner, 1998; Chen et al., 2000) and satellite photometry (Clancy and Rush, 1989; Clancy et al.,1994; Leblanc and Hauchecorne, 1997). Various characteristics of the MIL at low latitude (Kumar et al., 2001; Nee et al., 2002; Ratnam et al., 2003; Fadnavis and Beig, 2004) and mid latitudes (Leblanc and Hauchecorne 1997; Meriwether et al., 1998; States and Gardner 1998; Thomas et al., 2001) have been studied by exploring variations of the amplitude, frequency of occurrence, height and width of the MIL.

The physical mechanism producing and sustaining the inversion layer is not well understood; different mechanisms have been proposed from time to time to explain the observed characteristics. States and Gardner (1998) attributed the MIL to diurnal tides. Chu et al. (2005), using night time lidar measurements from the Starfire Optical Range (SOR) of New Mexico and Maui (Hawaii), attributed the MIL observed between 85-101 km to diurnal or semidiurnal tides. Further, the coupling of gravity waves (GWs) and the mesopause tidal structure was also proposed as the cause of MIL (Meriwether et al., 1998; Liu and Hagan, 1998; Liu et al., 2000). Mlynczak and Solomon (1991, 1993) and Meriwether and Mlynczak (1995) explored the possibility of chemical heating of the mesosphere by exothermic reactions. Hauchecorne et al. (1987), Senft and Gardner (1991), Meriwether et al. (1994), Whiteway et al. (1995), Thomas et al. (1996), Sica and Thorslay (1996), Gardner and Yang (1998) and Meriwether et al. (1998) indicated that GW breaking plays an important role in the development of MIL. Various processes such as tidal disturbances, chemical heating, tides/gravity waves-mean flow interactions, semiannual oscillations and inertial instabilities may either singly or collectively be operative in the production of the MIL.

Two-dimensional modeling studies reveal that GWs play a significant role in amplifying the temperature amplitude of the tidal structure and thus contribute to strong MILs (Hauchecorne and Maillard, 1990; Leblanc et al., 1995). Hauchecorne et al. (1987) described a model in which a succession of breaking GWs would generate a MIL through the gradual accumulation of heat. The breaking and dissipation of gravity waves provide a feedback mechanism causing turbulent heating which maintains the MIL (Meriwether and Gardner, 2000). Upward propagating gravity waves produce turbulence and turbulent viscous type mixing becomes important for tidal dissipation in the mesosphere and lower thermosphere (Akmaev, 2001 a, b). Akmaev (2001b) simulated the large-scale dynamics of the mesosphere and lower thermosphere with parameterized gravity waves and produced realistic simulations of the zonal mean thermal structure and winds and their seasonal variations in the mesosphere and lower thermosphere. The simulation results show that the diurnal tidal variation is a leading contributor to the semi-diurnal variation in the occurrence of inversion layers at low latitudes (Akmaev, 2001a,b, and references therein). However, Fadnavis and Beig (2004) emphasized that chemical heating involving ozone reactions may be a possible contributor to the production of MIL. Detailed discussions on the possible mechanisms for MIL production are given by Meriwether and Gardner (2000).

In the real atmosphere, thunderstorm updrafts and tropospheric convection cause the generation of internal gravity waves which penetrate up into the middle atmosphere. The dissipation of gravity waves may result either in heating or cooling depending upon the prevailing conditions in the atmosphere. In the upper middle atmosphere these waves may dissipate due to a local convective instability and may produce heating near 70 km altitude (Goya and Miyahara, 1999). Several papers report evidence of the relationship between observed gravity waves and strong convective activity (Erickson et al., 1973, Allen and Vincent, 1995; Walterscheid et al., 2001; Alexander et al., 2005,). The propagation of gravity waves from the troposphere to the mesosphere is strongly influenced by the zonal wind which in turn is related to the global zonal and meridional circulation in the middle atmosphere ( Delisi and Dunkerton , 1988). Thus the semi – annual oscillation in gravity wave activity may be related with zonal wind profile. Garcia and Solomon (1985) explained the semi-annual oscillations in mesospheric ozone due to filtering of gravity waves by the equatorial wind profile , modulating the transport of $H_2O$ and the associated destruction of ozone

In this paper we study a possible relationship between thunderstorms and mesospheric heating near 70-85 km (as evaluated by MIL). For this purpose, we have analyzed the height profile of temperature, and mesospheric ozone volume mixing ratio, for the period October 1991- September 2001 and the thunderstorm occurrence rate for the period April 1995-March 2000. The sources of the data used and its analysis are described in the next section-2. The results are discussed in section 3. The proposed mechanism and its interpretation are discussed in section 4. Although, due to the limitation of the availability of data, we have limited our analysis to different regions of India, the results of the analysis should be applicable to other regions as well. Finally, some conclusions are given in section 5.

## 2. Data and Analysis

The vertical structure of the middle atmospheric temperature and ozone volume mixing ratio along with other parameters have been regularly measured by the Halogen Occultation Experiment (HALOE) aboard the Upper Atmospheric Research Satellite (UARS) since October 1991. These profiles are interpolated onto a standard set of pressure levels in the altitude range between 10 and 130 km, with a vertical resolution of about 2.3 km between pressure levels. Since HALOE is a solar occultation instrument, solar infrared measurements are only made during UARS orbit sunrises and sunsets. Latitudinal coverage is from 80$^{o}$S to 80$^{o}$N over the course of a year. The UARS orbit has an inclination of 57$^{0}$ and a period of about 96 minutes. This results in the measurement of thirty profiles per day at two quasi-fixed latitudes, that is 15 at one latitude, corresponding to sunrise, and 15 at another corresponding to sunset.

The HALOE (Version 19, level 3AT) temperature results are for the period October 1991 to September 2001 over the altitude range of 34-86 km; above 86 km, HALOE uses MSISE-90 model results, which are used in the present study. The error in the temperature varies from 1K at 30 km to ~20 K at 90 km (10 K at 75 km). When daily temperature profiles from 50 km to 100 km were plotted, the most notable feature in the temperature structure of mesosphere, the inversions, could be observed. Standard vertical temperature profiles show a decrease in temperature above the stratopause with increasing altitude, up to the mesopause, above which the temperature increases with altitude. These are non inversion days, as shown in figure 1(a). Figure 1(b) shows that the temperature decreases above the stratopause as usual, but at about 75 km it starts to increase up to 77-83 km and then it again decreases to ~90km; these specific days are called mesospheric inversion days. The altitude range ~75-85 km in the mesosphere where temperature reversal takes place is known as the inversion layer. The difference between the maximum temperature and minimum temperature in this range is known as the amplitude of inversion layer. The altitude where the temperature becomes a minimum and then starts increasing is known as the bottom of the inversion layer. The altitude where the temperature becomes a maximum and then again decreases is known as the top (peak) of the inversion layer. The thickness is the difference between the altitude at the top and the bottom of the inversion layer.

Details of the calculation of the amplitude and occurrence rate of the MIL are given by Fadnavis and Beig (2004). We have considered four geographic regions: (1) Indian region (0-30$^0$N, 60-100$^0$E), which is further subdivided into two latitudinal bands (2) Band-1 (0-15$^0$N, 60-100$^0$E) and (3) Band-2 (16-30$^0$N, 60-100$^0$E), and (4) the oceanic region (17.5$^0$S-2.6$^0$S; 56.5$^0$E-71.5$^0$E). Daily temperature profiles are analyzed for these regions to study the frequency of occurrence and amplitude of the MIL. The amplitude of the inversion layer, if present, is obtained on a daily basis; these values are averaged to obtain monthly mean values. These monthly means are further averaged for each month over 10 year period to study seasonal variations over all the regions. Because errors in temperature estimates near the altitudes of the MIL are about 10K, inversions of 10K amplitude or more can be considered as being significant. The amplitude of the inversion varies between ~4 and 24K, with an average amplitude of around 12K being observed over all the three regions (1) to (3).

Ozone volume mixing ratios (VMRs) obtained from HALOE are extracted over these Indian regions for October 1991-September 2001 and over the ocean region for April 1995-March 2000. The ozone VMRs are averaged over the altitude range from 60-70 km near the bottom of the MIL, from 70-85 km where the MIL occurs, and from 85-90 km near the warm top of the MIL, in order to study the characteristic variations of ozone below, within and on top of the MIL.

Thunderstorm related information is derived from Optical Transient Detector (OTD) data, which is a scientific payload on the MicroLab-1 satellite, launched in April 1995. The spatial resolution of the instrument is 10 km and the temporal resolution is 2 ms. The OTD instrument detects both intra-cloud and cloud-to-ground discharges during day and night conditions, with a high detection efficiency. The orbital trajectory of the MicroLab-1 satellite allows the OTD to circle the Earth once every 100 minutes at an altitude of 740 km. Using its 128 x 128 pixel photo-diode array and wide field-of-view lens, the OTD instrument views an area of 1300 km x 1300 km. Given the field-of-view and the orbital trajectory, the OTD can monitor individual storms and storm systems for about 4 minutes (Christian et al., 2003). OTD data is available till March 2000. The number of flashes recorded by OTD over the period April 1995 - March 2000 is obtained over the Indian region (0-30$^0$N, 60-100$^0$E), the band-1 (0-15$^0$N, 60-100$^0$E), the band-2 (16-30$^0$N, 60-100$^0$E) and over the ocean (17.5$^0$S to 2.6$^0$S; 56.5$^0$E to 71.5$^0$E) from the website:

http://thunder.nsstc.nasa.gov/lightning-cgi-bin/otd/OTDSearch.pl

The total number of flashes for a month is summed over the 5 year period to obtain the monthly distributions for the geographic regions.

Ground based thunderstorm data for the period April 1995 – March 2000 are obtained from 78 Indian observatories spread over the Indian region (0-30$^0$ N, 60-100$^0$E), the band-1(0-15$^0$N, 60-100$^0$E) and the band-2(16-30$^0$N, 60-100$^0$E). These data have been extracted from summaries published by the Indian Meteorological Department (IMD, 1995-2000).

To study the possible linkage between the MIL and the lightning/thunderstorm occurrence, the monthly amplitude of the MIL and the frequency of occurrence are also considered for the operational period of the OTD.

## 3. Results and Discussions

Figure 1(a) and 1(b) exhibit typical vertical temperature structures for randomly selected non-inversion days (28 April, 21 July, 22 July 1995) and inversion days (9 May, 2 November and 4 November 1995), respectively; successive temperature profiles are shifted by 20K. On 9 May 1995, a minimum temperature of 186K is observed at an altitude of 72 km above which it does not follow the usual fall in temperature as per the lapse rate. Instead it increases with height and reaches a maximum of 209 K at an altitude of 80 km, above which it again decreases. The amplitude of the inversion for this layer is thus found to be ~23 K the inversion layer can be seen in other profiles, which are marked by double-pointed arrows in Figure 1. All the profiles are examined for the entire 10 years and the amplitude, and occurrence, of the mesospheric inversion layer are obtained for each day. These results clearly indicate that on inversion days the temperature increases by ~20 K within 70-85 km altitude range.

The monthly mean amplitude of the inversion temperature is obtained over all three continental regions. Figure 2 displays the time series of the monthly mean variation of the amplitude of the inversion over the Indian region, the band-1 and the band-2. The amplitude of the inversion varies from 4 to 16K over the Indian region, from 4 to 24 K over the band-1 and from 4 to 20 K over the band-2. The amplitude of the inversion is on average, ~12 K over all three regions. Amplitude of inversion has been calculated from the data extracted over the Indian region (0-30$^0$N, 60-100$^0$E), the band-1 (0-15$^0$N, 60-100$^0$E) and the band-2 (16-30$^0$N, 60-100$^0$E) separately on a daily basis. Temperature profiles over the Indian regions include daily profiles of band-1 as well as those of band-

2. Amplitudes obtained from these profiles are then averaged for a month for all the years for the Indian region. Hence amplitude gets averaged. This averaged amplitude appears over the Indian region. For example in October 1992 amplitude of inversion over the band-1 was 12.58K and over the band-2 was 9.73K. If we average these amplitudes we get ((12.58+9.73)/2.0)= 11.15K, which is close to observed amplitude over the Indian region (~10.5K). Hence amplitude of two individual Indian sub-regions reaches higher values (4 K to 24 K over the band-1 and 4 to 20 K over the band-2) than the Indian region (4K to 16K).

To study seasonal variations the amplitude of the inversions for each month is averaged for the entire 10 year period. Figure 3(a) and 3(b) exhibit the seasonal variation of frequency of occurrence and amplitude of the MIL along with two sigma error bars for the three regions. Both the amplitude and the frequency of occurrence show a semiannual variation over all regions. Over the band-1, the maxima occur in April (just after spring equinox) and November (just after fall equinox). The amplitude of the temperature inversion shows maxima in the months of April or May and October or November. The monthly mean amplitude varies from 7 to 12 K over the Indian region, from 8 to 16 K over the band-1, and from 5 to 13K over the band-2, in good agreement with values reported by Leblanc and Hauchecorne (1997), who obtained the seasonal mean amplitude of the temperature inversion for the period September 1991-August 1995 to vary from 2 K to 10K. The strong semiannual variation with maxima one month after the equinoxes has also been reported by Fadnavis and Beig, (2004, and references therein).

In order to study the relationship between both the frequency of occurrence and the amplitude of the MIL and thunderstorm activity, data are analyzed for the period

April 1995-March 2000, when OTD data were available. Figure 4(a) and 4(b) exhibit the seasonal variation of frequency of occurrence and MIL amplitude along with two sigma error bars for 5 year period. They are rather similar to results shown in Figure 3. Figure 5(a) shows the seasonal variation of frequency of occurrence and (b) the MIL amplitude with the number of lightning flashes as obtained from OTD over the respective regions for the same periods. The annual variations for the two parameters are the same, with maxima during the equinoxes and minima during summer and winter for all three regions. In all the figures we have shown the correlation coefficient between the variables (left and right vertical axes), which lies between 0.50 and 0.72.

Figure 6 (a) and 6(b) shows seasonal variation of frequency of occurrence and MIL amplitude versus the number of thunderstorm days obtained from widely spread stations over the respective regions. It is interesting to note that the annual variation of the number of thunderstorm days shows characteristic variation with variation of latitude. Closest to the equator (0-15$^0$N, 60-100$^0$E) the maximum is observed during May and October, with the May maximum being stronger than the October maximum. For the Band-2 (16-30$^0$N, 60-100$^0$E) the maximum number of thunderstorm days occurs during May and September, with the September maximum being stronger. Manohar et al. (1999) have reported the latitudinal distribution of thunderstorm-activity over the Indian region using these station data for the period 1970-1980. The latitudinal variation of the mean number of thunderstorm days indicates systematic changes in the semiannual oscillation, which are consistent with the migration of the Inter-Tropical Convergence Zone (ITCZ) over different latitudinal belts. Over the band-1 the first maximum occurs in April-May and the second in October-November; the first maximum is stronger than the second. In the case of the band-2, the first maximum is in May whereas the second is in September, and is stronger than the May one (Manohar et al., 1999). Our results show similar variations. Frequency of occurrence and amplitude of inversion, exhibit fairly good correlation, with the number of thunderstorm days over the Indian region. The correlation is poorer over the band-1 and the band-2. The reason for this may be that numbers of land stations are smaller there. Over the entire tropical region (uppermost panels) the frequency of occurrence and the MIL amplitude exhibit fairly good correlations (>0.5) as there are larger number of stations.

The number of lightning flashes as obtained from OTD data shows a better correlation with the frequency of occurrence and the MIL amplitude over all three regions (Figure 5) than with the number of thunder storm day (Figure 6). The reason may be that both HALOE and OTD view the sky from above and cover both land and ocean regions. Lightning flashes and number of thunderstorms days show better correlation with frequency of occurrence than the MIL amplitude. This indicates that inversion may be occurring with thunderstorm activity but its amplitude may not be varying proportionately.

Figure 7 (a) shows the seasonal variation of frequency of occurrence of the mesospheric temperature inversion and the number of lightning flashes recorded by OTD satellite for the period April 1995 -March 2000 over the open ocean South of India ($2.5^0$S - $17.5^0$S; $56.5^0$E - $71.5^0$E). Figure 7(b) shows the seasonal distribution of amplitude of temperature inversion versus the number of lightning flashes recorded by the OTD satellite for the same period and over the same area. It is evident that the number of lightning flashes and the frequency of MIL occurrence (%) are considerably less over the ocean than over to land, but the amplitude of the temperature inversions are of the same magnitude (see Figure 5). Both the frequency of occurrence (r = 0.53) and the amplitude of the mesospheric temperature inversion (r = 0.65) show fairly good correlation with the number of lightning flashes, and both have a strong semiannual variation.

To study the characteristic variation of ozone near the cold bottom of the MIL, within the MIL and near the warm top of the MIL, ozone volume mixing ratios over the Indian tropical region for the period October 1991- September 2001 have been averaged over 65-70 km, 70-85 km and 85-90 km, respectively. To smooth the time series, six-point running averages were obtained. Figure 8 indicates the time series of the average ozone volume mixing ratio (scatter plot) and the six-point running averages (thick line), near the bottom of the MIL (65-70 km), within the MIL region (70-85 km) and at the top of the MIL (85-90 km) region. Semiannual oscillations in ozone content are quite evident for all three altitudes, but with a strong annual oscillation at 65-70km. It is interesting to note that the ozone content reverses its phase in region of 70-85 km, where the MIL is most frequently observed. The ozone volume mixing ratio at 70 – 85km is ~0.4 times its value observed at 65-70 km.

To study the relation between the frequency of occurrence and the amplitude of the temperature inversion with the variation of ozone content, the frequency of occurrence and the amplitude of the temperature inversion calculated for the period October 1991-September 2001 are averaged for each month. Ozone volume mixing ratios are averaged over the altitude range (70-85 km) where the MIL is most frequently observed for each month. Figure 9 (a) and 9(b) show the seasonal variation of frequency of occurrence and the MIL amplitude together with the average ozone volume mixing ratio (Part per billon by volume). Ozone amounts show a semiannual variation. Both the frequency of occurrence and the amplitude of the MIL show a good correlation with the seasonal variation of ozone.

Over the ocean region the inversions are observed in the 70-80 km altitude range; hence, the ozone volume mixing ratios are averaged over 70-80 km. Figure 10 indicates the seasonal variation of frequency of occurrence and MIL amplitude with the average ozone volume mixing ratios (ppb) over the oceanic region for the period April 1995 - March 2000. Again ozone amounts show a semiannual variation. The frequency of occurrence and the amplitude of the mesospheric temperature inversion show correlation coefficients of 0.31 and 0.12 respectively; these are statistically insignificant at the 99% level. Because the ozone concentrations are somewhat less over the ocean region than over the land, the chemical heating in the mesosphere will be less. This may be a reason for the insignificant correlation between the variation of ozone and the frequency of occurrence and the MIL amplitude over this oceanic region.

The semiannual oscillation in ozone amounts for the altitude range 70-90 Km over the tropics was reported from satellite data (Richard et al., 1990; Thomas, 1990). The ozone volume mixing ratios show the maximum values during April and November over all three Indian regions. From Solar Mesosphere Explorer (SME) satellite measurements for the period 1982-1983, Thomas and Barth (1984) reported that the ozone density near 80 km shows large seasonal changes. During the equinoxes the ozone densities are about 2-3 times those observed at the solstices, as is also evident in the present study. Thomas and Barth (1984) proposed that seasonal variability in ozone may be understood in terms of transport by gravity waves in the mesosphere, which in turn result from the seasonal modulation of the propagation and breaking of small scale gravity waves there.

4. Mechanism and Interpretations

In the above figures, we have shown that over India, semiannual variation in the frequency of occurrence of mesospheric temperature inversion layer has a fairly good correlation with semiannual variations in thunderstorm activity and averaged ozone volume mixing ratios. Both the frequency of occurrence and the amplitude of the temperature inversion layer show maximum values during April-May (in the pre monsoon season) and October-November, when thunderstorm activity is also high. Thunderstorm activity over India shows both a latitudinal and a seasonal variation in the pre-monsoon season (Manohar et al., 1999). The seasonal variation is related to the migration of the monsoon trough or Inter-Tropical Conversion Zone (ITCZ) from the band-1 (0-15$^0$N, 60-100$^0$E) to band - 2 (16-30$^0$N, 60-100$^0$E). The band-1 shows maximum thunderstorm activity during April, as the monsoon trough marches northward; the band-2 shows maximum thunderstorm activity during May. During June- July the monsoon spreads all over India and therefore, the number of thunderstorms observed is less at this time. The withdrawal of monsoon starts in the northern India and progresses towards the South, thunderstorm activity increases; during September thunderstorm activity is high over band-2 and during October over band-1.

The formation of the MIL has been simulated in numerical models of the atmosphere, including the local drag on the zonal wind associated with gravity wave breaking capable of creating local adiabatic circulation and hence adiabatic heating and cooling (Hauchecorne and Maillard, 1990; Leblanc et al., 1995). Liu and Gardner (2005) tried to explain the observed weaker MIL at Maui than at SOR, New Mexico in terms of weaker gravity wave dissipation and therefore a weaker downward flux of atomic oxygen, which is associated with several chemical reactions in this region that converts solar energy to the chemical heating leading to the MIL. This shows that gravity wave breaking and dissipation play significant roles in the formation/existence of MIL.

HALOE and OTD being polar orbiting satellites, their simultaneous passes over a specific region are very few. Table 1 and Table 2 indicate typical days where simultaneous HALOE, OTD and ground based station data are available over a common region. Table 1 indicates typical non inversion days when neither thunderstorm activity was reported at nearby ground stations nor lightning flashes recorded by OTD in a $3^0$ x $3^0$ grid centered at the position of HALOE. Table 2 shows typical inversion days when thunderstorm activity was reported at nearby ground stations and lightning flashes were recorded by OTD.

The rapid and deep convection associated with thunder storm generated gravity waves which propagate upwards and may get amplified. The gravity waves become unstable at the height where the zonal wind velocity becomes equal to the wave phase velocity. Usually in mesosphere this condition is satisfied and breaking of gravity waves takes place (Sica and Thorsley, 1996: Thomas et al., 1996). The turbulent heating, arising from the breaking of waves, provides a feedback mechanism that then may maintain the observed MIL. A strong seasonal dependence of mesospheric GW activity is observed, with a peak in the summer months and much reduced activity during the winter months over a mid-latitudinal station in Michigan, U.S. A. ($42.3^0$ N, 83.7 $^0$W) (Wu and Killeen, 1996).

Temperature profiles obtained from ground based MST radars and Lidars show fluctuations in nighttime temperatures with characteristic scales resembling those of large-scale gravity waves (Parameswaran et al., 2000). The dynamics of dissipating GWs will transport heat downward. Thus, the transport of heat below 80 km by dissipating GWs may contribute to the formation of the MIL near 70 km (Meriwether and Gardner, 2000). A two dimensional model simulation of gravity wave-tidal coupling points strongly points to the conclusion that the breaking of gravity waves may play a significant role by amplifying the temperature amplitude of the tidal structure and producing a large MIL (Liu and Hagan, 1998; Liu et al., 2000). Although it is known that the tidal variation is the significant contributor to the semidiurnal variation in the occurrence of inversion layers at low latitudes, we could not evaluate the contribution of tides from the HALOE data, due to its poor sampling in local time.

Heating due to several exothermic chemical reactions (implying ozone) could also promote temperature inversions (Meriwether and Mlynczak, 1995; Mlynczak and Solomon, 1993; Fadnavis and Beig, 2004). The seasonal modulation of the propagation and breaking of small-scale gravity waves leads to the variation of gravity-wave-induced transport in the mesosphere, which produces seasonal variability in ozone (Thomas and Barth, 1984). Garcia and Solomon (1985) have discussed that the filtering of gravity waves by the equatorial wind profile is responsible for modulating the transport of $H_2O$ and associated destruction of ozone. The gravity waves produced during thunderstorms may contribute to the production of MIL, and may perturb the semiannual oscillation (reverse phases observed in the MIL). Gravity waves may also lead to the transport of atomic oxygen and through the route of chemical heating may contribute to the production of ozone. This established correlation between thunderstorm, MIL and ozone transport.

Sentman et al. (2003) reported the simultaneous observation of coincident gravity waves and sprites in the mesosphere, emanating from the same underlying thunderstorm. Sprites are the results of atmospheric heating by the electromagnetic pulse generated by a powerful lightning stroke (Füllekrug et al., 2006, Barrington-Leigh and Inan, 1999, Siingh et al., 2005, 2007, and references therein). Sprites may also change the concentration of $NO_x$ and $HO_x$ in the mesosphere (Hiraki et al., 2002). These chemical changes may have an impact on the observed cooling/heating in the middle atmosphere, which may influence MIL. This aspect has not yet been explored. Attempt should be made to find out if there is any relation between MILs and sprites.

## 5. Conclusions

The seasonal variation of the frequency of occurrence and the amplitude of the mesospheric inversion layer obtained from HALOE temperature profiles over the entire Indian region (0-30$^0$ N, 60-100$^0$E), the band-1(0-15$^0$N, 60-100$^0$E), the band-2 (16-30$^0$N, 60-100$^0$E) and the oceanic region (17.5$^0$S - 2.5$^0$S; 56.5$^0$E - 71.5$^0$E) show strong semiannual variations. Both the frequency of occurrence and the amplitude of the temperature inversion exhibit a maximum and minimum in the same month over the respective regions except for the Indian region. The semiannual oscillation in ozone below the cold bottom of the MIL reverses its phase within the MIL region. Above the warm top of the MIL, the phase of the semiannual oscillation does not show such a reversal but its amplitude increases. The Seasonal variation of the frequency of occurrence and amplitude of the inversion exhibit a fairly good correlation with the seasonal variation of the number of lightning flashes and of thunderstorm days. The seasonal variation of frequency of the occurrence and the amplitude of the inversion also exhibits a fairly good correlation with the seasonal variation of ozone over these regions. The frequency of occurrence of the MIL and thunderstorm activity are less over the oceanic region than over the land. Thunderstorm activity is observed by OTD and nearby ground station data on strong inversion days, and no thunderstorm activity is observed by OTD and nearby ground station data on non inversion days. This all indicates that gravity waves produced during thunderstorms, together with chemical heating due to ozone, may contribute significantly in the production of the mesospheric inversion layer.

Acknowledgements

We (S.F. and G.B.) acknowledge the Climate And Weather of Sun Earth System–India (CAWSES) program of the Indian Space Research Organization for financial assistance to this project. D. Siingh would like to acknowledge the DST, under the BOYSCAST programme (reference SR/BY/A-19/05). We are also grateful to anonymous reviewers for their valuable suggestions. We thank reviewer II for his help in English language corrections.

**Table and Figure Captions**:

Table 1: Non inversion days observed in the HALOE temperature profile and no thunderstorm activity reported at a nearby station and no lightning flashes recorded by the OTD in this vicinity on the same day.

Table 2: Inversion days observed in the HALOE temperature profile and thunderstorm activity reported at a nearby station and the number of lightning flashes recorded by the OTD on the same day.

Figure1: Vertical temperature structure for the three randomly selected typical (a) non inversion days (28 April, 21 July, 22 July 1995) and (b) inversion days (9 May, 2 November, 4 November, 1995) within the Indian tropical belt recorded by HALOE. The interval between the double-pointed arrows represents the width of the inversion layer.

Figure 2. Monthly variation of the amplitude of the temperature inversion derived from the HALOE temperature series over the Indian region (0 -30$^0$ N, 60 -100$^0$ E), the band-1(0 - 15$^0$N, 60 - 100$^0$E) and the band-2 (16-30$^0$ N, 60-100$^0$ E) for the period 1991- 2001.

Figure 3. Seasonal variations of (a) Frequency of occurrence of MIL, (b) Amplitude of MIL derived from the HALOE temperature series over the Indian region (0-30$^0$ N, 60-100$^0$ E) for the period 1991- 2001. The error bars show ±2 sigma values.

Figure 4. Seasonal variation of (a) Frequency of occurrence of MIL, (b) Amplitude of MIL using the HALOE temperature series over the Indian region (0-30$^0$ N, 60-100$^0$ E) for the period April 1995 - March 2000. The error bars show ±2 sigma values.

Figure 5. Seasonal variation of the (a) Frequency of occurrence of temperature

inversion (%) obtained from the HALOE temperature data and number of lightning flashes as obtained from OTD data for the period April 1995 - March 2000, over the Indian region (0 - 30$^0$ N, Long. 60 -100$^0$ E), the band-1 (0-15$^0$ N, 60-100$^0$ E) and the band-2 (16 - 30$^0$N, 60 - 100$^0$ E). (b) Amplitude of MIL as obtained from the HALOE temperature data and number of lightning flashes as obtained from OTD data.

Figure 6. Seasonal variation of the (a) Frequency of occurrence of the temperature inversion (HALOE data) and the number of thunderstorm days as obtained from different stations also for the period April 1995 -March 2000, over the Indian region (0-30$^0$ N, 60-100$^0$ E), the band-1 (0-15$^0$ N, 60-100$^0$ E) and the band-2 (16-30$^0$ N, 60-100$^0$ E). (b) Amplitude of MIL and number of thunderstorm days as obtained from ground stations for the same period and locations.

Figure 7. Seasonal variation of the (a) Frequency of occurrence of the temperature inversion (HALOE data) and the number of lightning flashes as obtained from the OTD satellite, for the period April 1995 -March 2000, over the ocean are (17.5 - 2.5$^0$S; 56.5 - 71.5$^0$E). (b) Amplitude of MIL and the number of lightning flashes (OTD data), for the same period and over the same ocean region.

Figure 8. Seasonal variation in averaged ozone concentration (ppb) at 65 -70 km, 70 – 85 km and 85 – 90 km over the Indian region (0 -30$^0$ N, 60 -100$^0$ E) for the period October 1991-September 2001 (scatter plot); the six points running average is also plotted (thick line).

Figure 9. Seasonal variation of the (a) Frequency of occurrence of the temperature inversion and the averaged ozone concentration at 70 – 85 km, both obtained from HALOE data for the period October 1991-September 2001, over the three Indian regions considered. (b) Amplitude of MIL and the averaged ozone concentration at 70 - 85 km, for the same period and location.

Figure 10. Seasonal variation of the (a) Frequency of occurrence of the temperature inversion and the averaged ozone concentration at 70 – 85 km from HALOE data for the period April 1995 -March 2000, over the ocean region (17.5 - 2.5$^0$S; 56.5 - 71.5$^0$E). (b) Amplitude of the MIL and the averaged ozone concentration at 70 - 85 km for the same period and location.

Table 1

| Non inversion day as observed in HALOE temperature profile (Day /Month/Year) | HALOE Position | No thunderstorm activity reported at nearby ground station | No lighting flashes recorded by OTD within $3^0$ x $3^0$ grid centered at |
|---|---|---|---|
| 21/07/95 | $12.83^0$ N, $83.56^0$ E | Vellore - $12.55^0$ N, $79.09^0$ E<br>Madras – $13.03^0$ N, $80.15^0$ E | ($12^0$ N, $80.5^0$ E) |
| 22/07/95 | $17.8^0$ N, $80.53^0$ E | Solapur – $17.4^0$ N, $75.54^0$ E<br>Hyderabad – $17.27^0$ N, $78.28^0$ E | ($17^0$ N, $78.5^0$ E) |
| 08/03/99 | $25.26^0$ N, $72.67^0$ E | Udaipur- $24.35^0$ N, $73.42^0$ E<br>Neemuch- $24.28^0$N, $74.54^0$E - | ( $25^0$ N, $72^0$ E) |
| 22/12/99 | $20.69^0$ N, $83.57^0$ E | Jagdalpur- $19.05^0$ N, $82.02^0$ E<br>Gopalpur- $19.16^0$ N, $84.53^0$ E | ($19^0$ N , $82^0$ E) |

Table 2

| Temperature inversion observed in HALOE temperature profile on (Day /Month/Year) | HALOE Position | Thunderstorm activity reported at ground station | Number of lighting flashes recorded by OTD within $3^0$ x $3^0$ grid centered at |
|---|---|---|---|
| 29/5/95 | $10.16^0$ N, $88.64^0$ E | Cuddalore- $11.46^0$ N, $79.45^0$ E<br>Nagapattinam-$10.46^0$ N, $79.5^0$ E | 30<br>($10^0$ N , $86^0$ E) |
| 27/7/97 | $22.66^0$ N, $71.05^0$ E | Ahemedabad- $23.04^0$ N, $72.38^0$ E<br>Ambikapur- $23.15^0$ N, $83.15^0$ E | 41<br>($22^0$ N, $72^0$ E) |
| 15/9/99 | $26.92^0$ N, $76.26^0$ E | Dubrugarh–$27.29^0$ N, $94.5^0$ E | 7<br>($26^0$ N $76^0$ E) |
| 5/2/2000 | $29.04^0$ N, $76.21^0$ E | New Delhi- $28.35^0$ N,$77.12^0$ E<br>Dubrugarh– $27.29^0$ N, $94.5^0$ E | 16<br>($27^0$ N, $77^0$ E) |

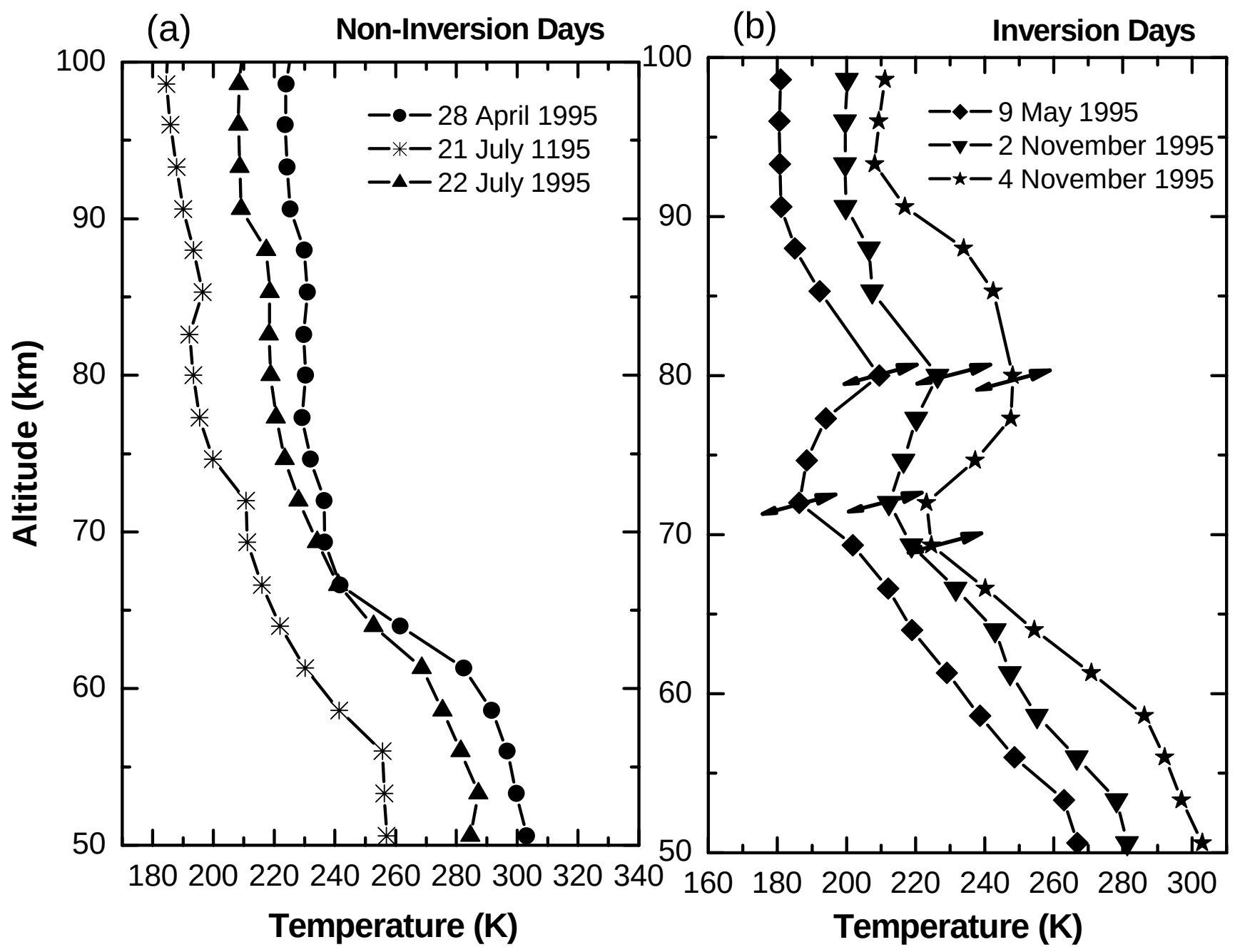

**Figure 1.**

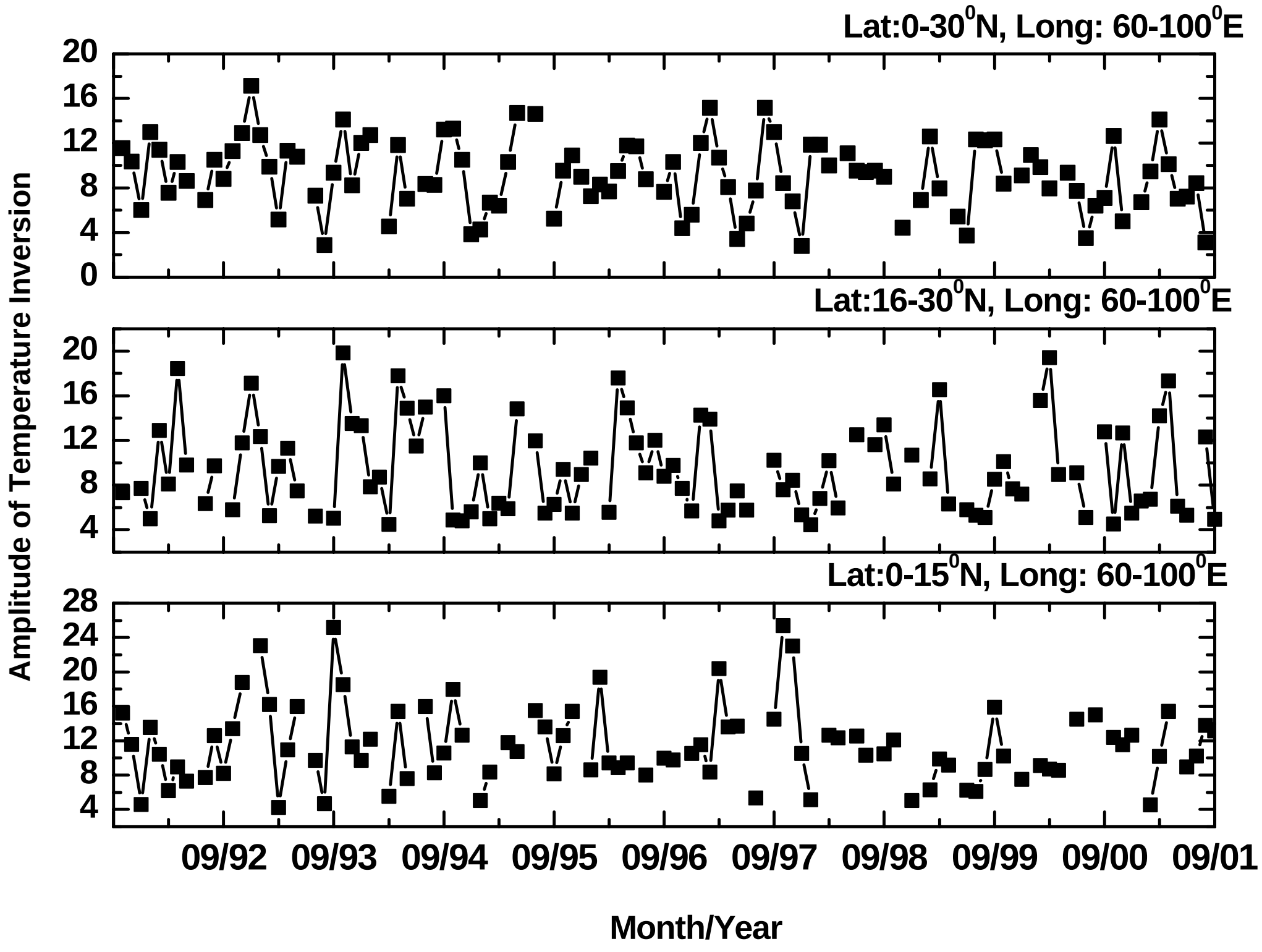

Figure 2.

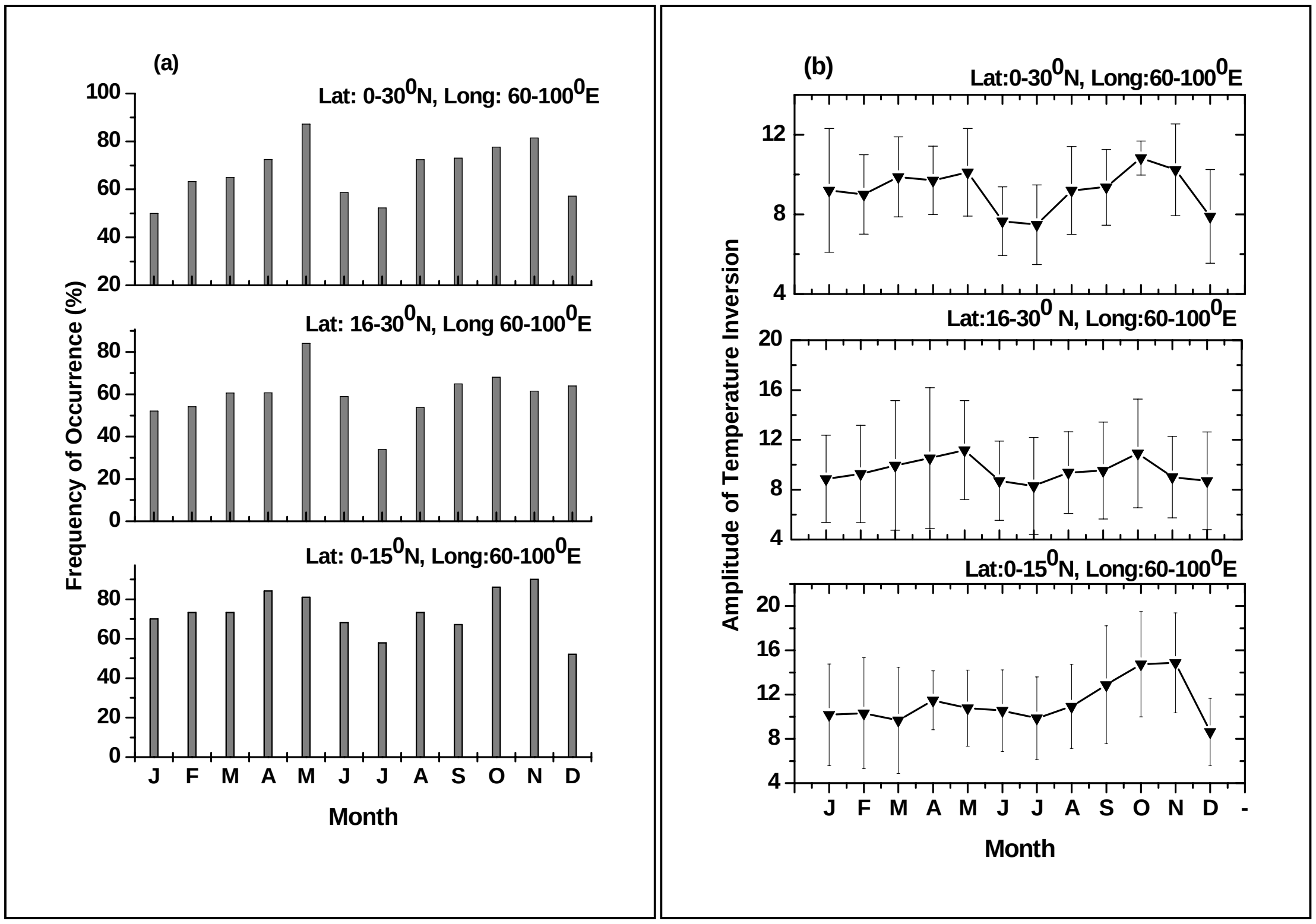
(a)
Lat: 0-30⁰N, Long: 60-100⁰E
Lat: 16-30⁰N, Long 60-100⁰E
Lat: 0-15⁰N, Long:60-100⁰E
Frequency of Occurrence (%)
J F M A M J J A S O N D
Month
(b)
Lat:0-30⁰N, Long:60-100⁰E
Lat:16-30⁰ N, Long:60-100⁰E
Lat:0-15⁰N, Long:60-100⁰E
Amplitude of Temperature Inversion
J F M A M J J A S O N D -
Month

Figure 3.

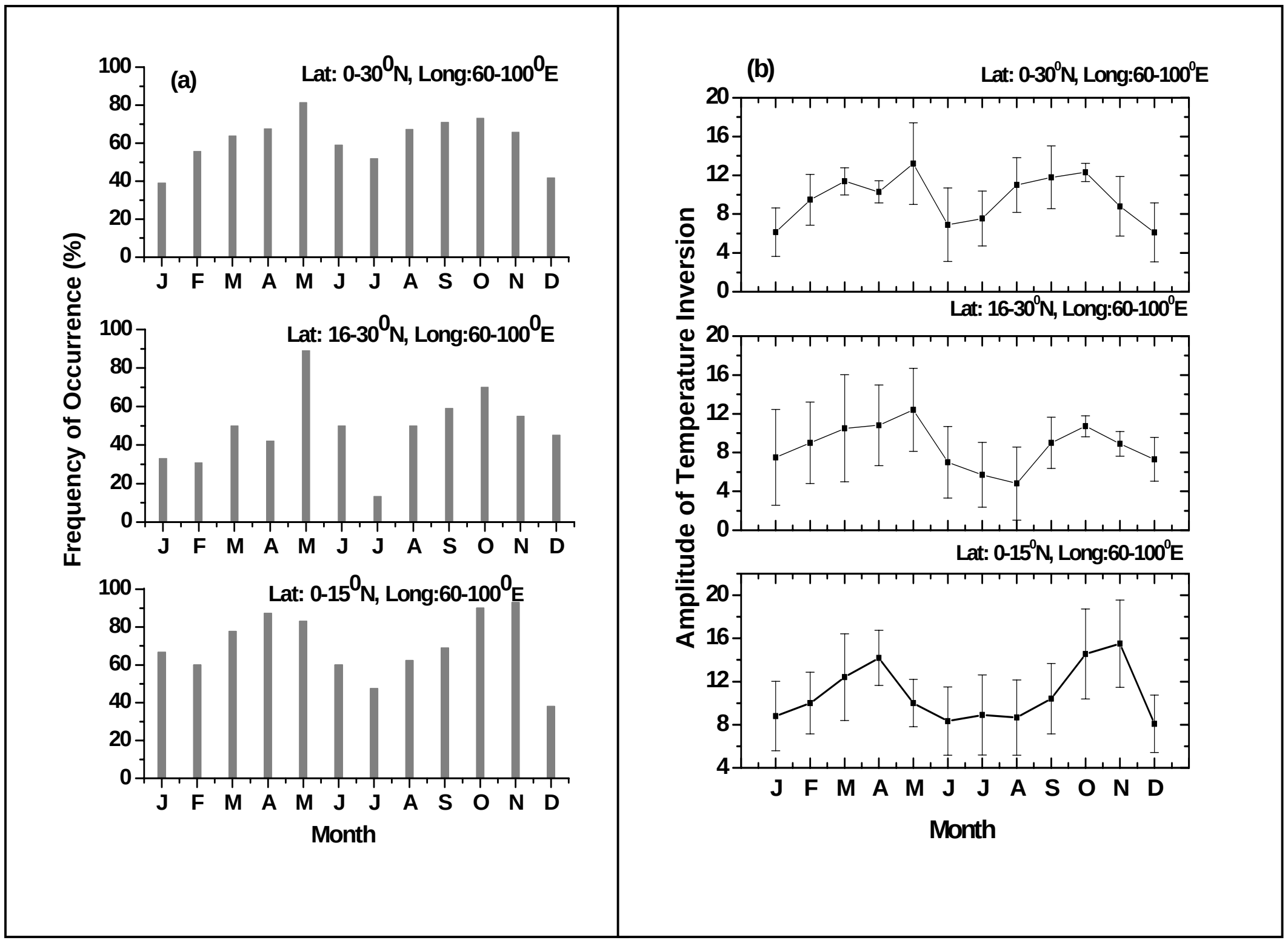

(a)
Lat: 0-30⁰N, Long:60-100⁰E
Lat: 16-30⁰N, Long:60-100⁰E
Lat: 0-15⁰N, Long:60-100⁰E
Frequency of Occurrence (%)
J F M A M J J A S O N D
Month
(b)
Lat: 0-30⁰N, Long:60-100⁰E
Lat: 16-30⁰N, Long:60-100⁰E
Lat: 0-15⁰N, Long:60-100⁰E
Amplitude of Temperature Inversion
J F M A M J J A S O N D
Month

Figure 4.

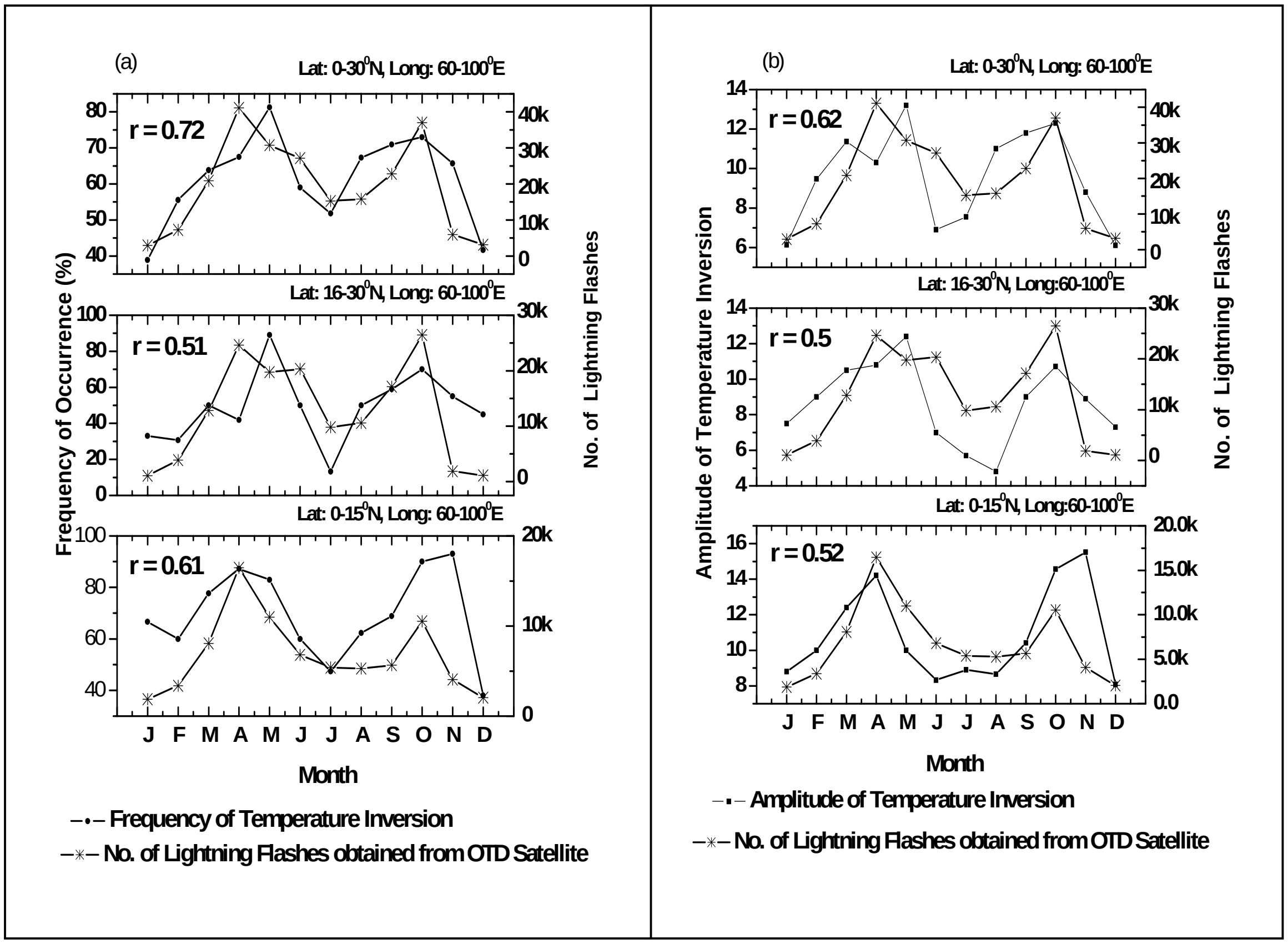
(a)
Lat: 0-30⁰N, Long: 60-100⁰E
r = 0.72
Lat: 16-30⁰N, Long: 60-100⁰E
r = 0.51
Lat: 0-15⁰N, Long: 60-100⁰E
r = 0.61
Frequency of Occurrence (%)
No. of Lightning Flashes
J F M A M J J A S O N D
Month
Frequency of Temperature Inversion
No. of Lightning Flashes obtained from OTD Satellite
(b)
Lat: 0-30⁰N, Long: 60-100⁰E
r = 0.62
Lat: 16-30⁰N, Long:60-100⁰E
r = 0.5
Lat: 0-15⁰N, Long:60-100⁰E
r = 0.52
Amplitude of Temperature Inversion
No. of Lightning Flashes
J F M A M J J A S O N D
Month
Amplitude of Temperature Inversion
No. of Lightning Flashes obtained from OTD Satellite

Figure 5.

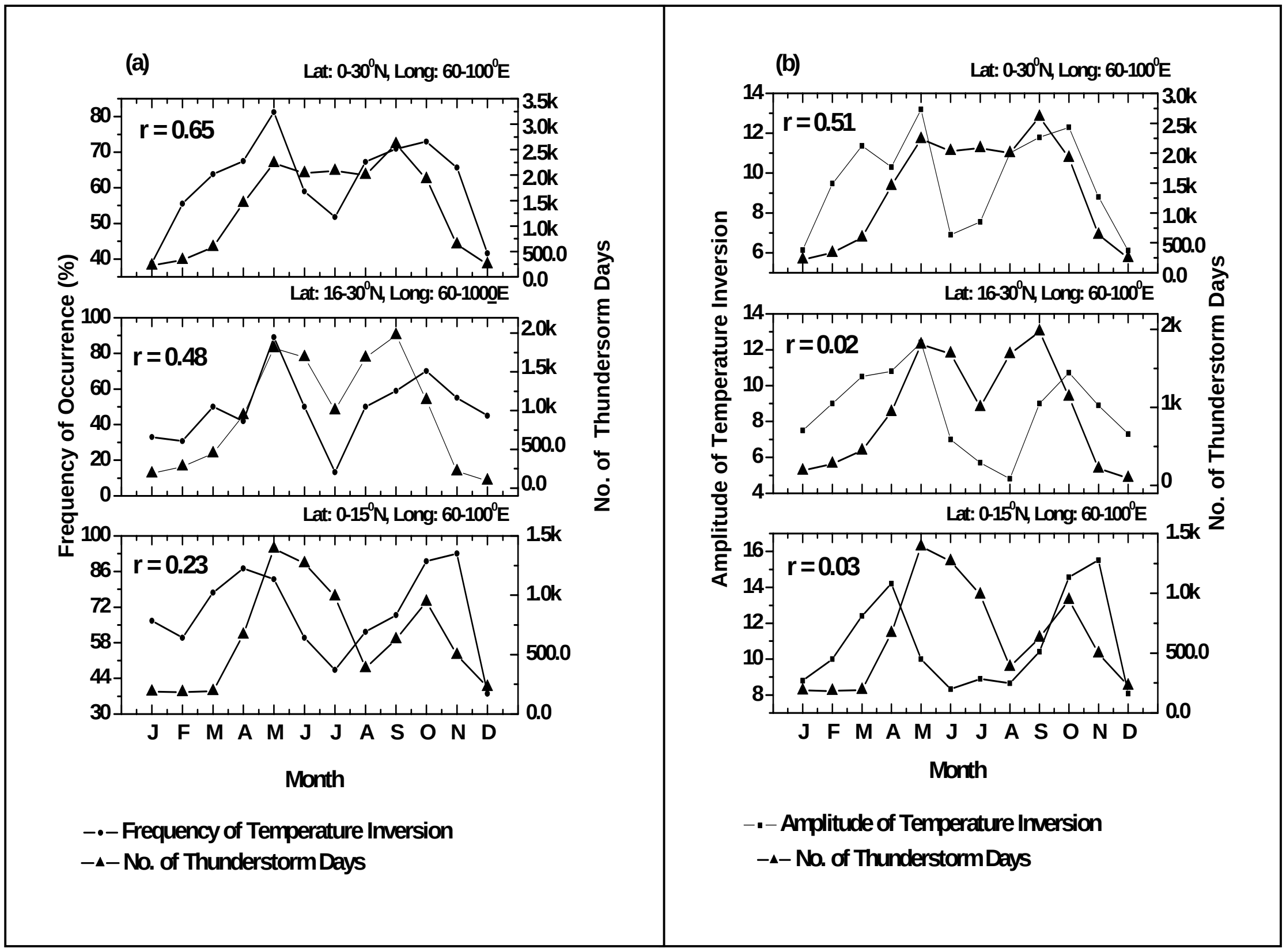
(a)
Lat: 0-30⁰N, Long: 60-100⁰E
r = 0.65
Lat: 16-30⁰N, Long: 60-1000E
r = 0.48
Lat: 0-15⁰N, Long: 60-100⁰E
r = 0.23
Frequency of Occurrence (%)
No. of Thundersorm Days
J F M A M J J A S O N D
Month
Frequency of Temperature Inversion
No. of Thunderstorm Days
(b)
Lat: 0-30⁰N, Long: 60-100⁰E
r = 0.51
Lat: 16-30⁰N, Long: 60-100⁰E
r = 0.02
Lat: 0-15⁰N, Long: 60-100⁰E
r = 0.03
Amplitude of Temperature Inversion
No. of Thunderstorm Days
J F M A M J J A S O N D
Month
Amplitude of Temperature Inversion
No. of Thunderstorm Days

**Figure 6.**

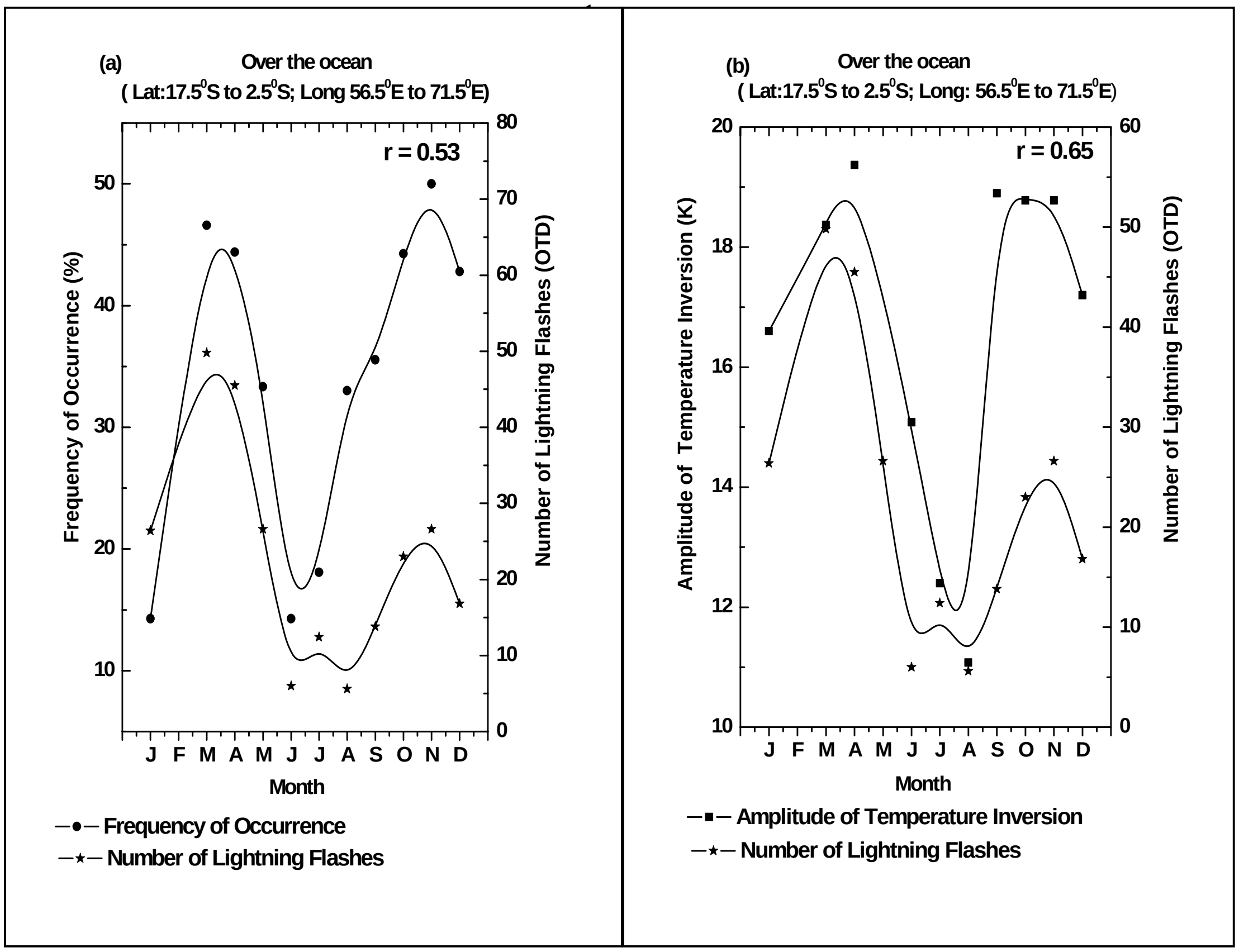
(a)
Over the ocean
( Lat:17.5⁰S to 2.5⁰S; Long 56.5⁰E to 71.5⁰E)
r = 0.53
Frequency of Occurrence (%)
Number of Lightning Flashes (OTD)
J F M A M J J A S O N D
Month
Frequency of Occurrence
Number of Lightning Flashes
(b)
Over the ocean
( Lat:17.5⁰S to 2.5⁰S; Long: 56.5⁰E to 71.5⁰E)
r = 0.65
Amplitude of Temperature Inversion (K)
Number of Lightning Flashes (OTD)
J F M A M J J A S O N D
Month
Amplitude of Temperature Inversion
Number of Lightning Flashes

Figure 7.

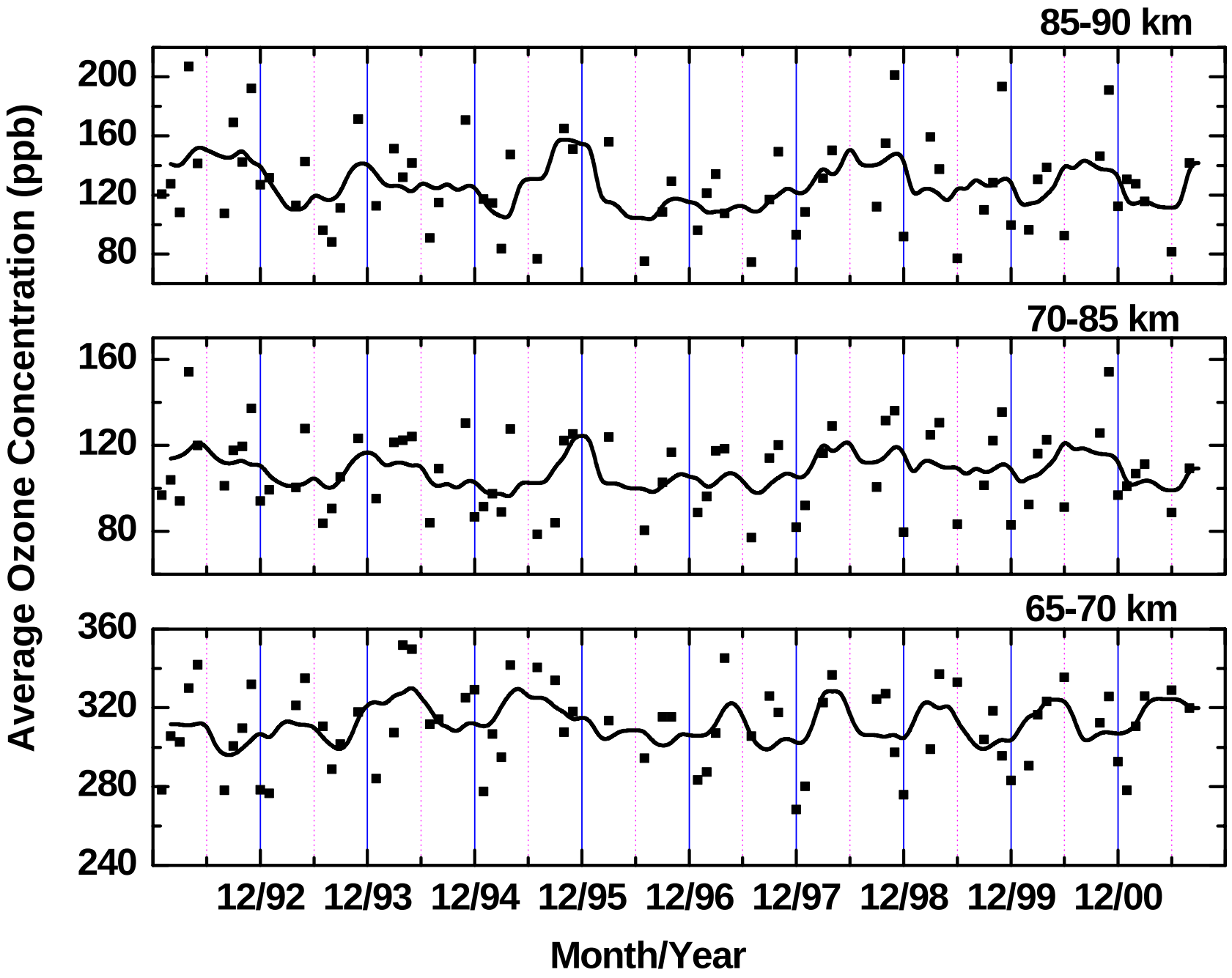

,,

Figure 8.

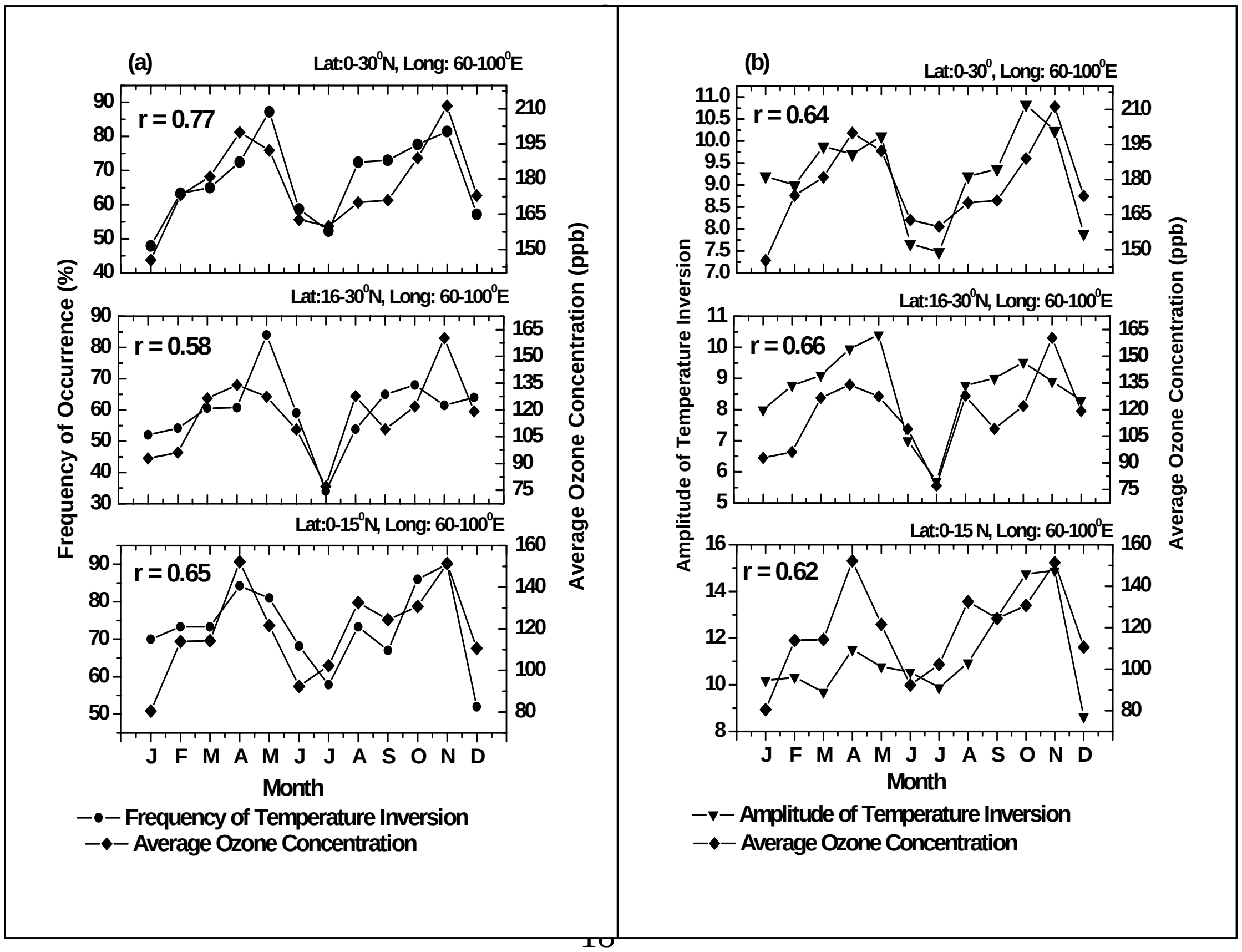
(a)
Lat:0-30⁰N, Long: 60-100⁰E
r = 0.77
Lat:16-30⁰N, Long: 60-100⁰E
r = 0.58
Lat:0-15⁰N, Long: 60-100⁰E
r = 0.65
Frequency of Occurrence (%)
Average Ozone Concentration (ppb)
J F M A M J J A S O N D
Month
Frequency of Temperature Inversion
Average Ozone Concentration
(b)
Lat:0-30⁰, Long: 60-100⁰E
r = 0.64
Lat:16-30⁰N, Long: 60-100⁰E
r = 0.66
Lat:0-15 N, Long: 60-100⁰E
r = 0.62
Amplitude of Temperature Inversion
Average Ozone Concentration (ppb)
J F M A M J J A S O N D
Month
Amplitude of Temperature Inversion
Average Ozone Concentration

**Figure 9.**

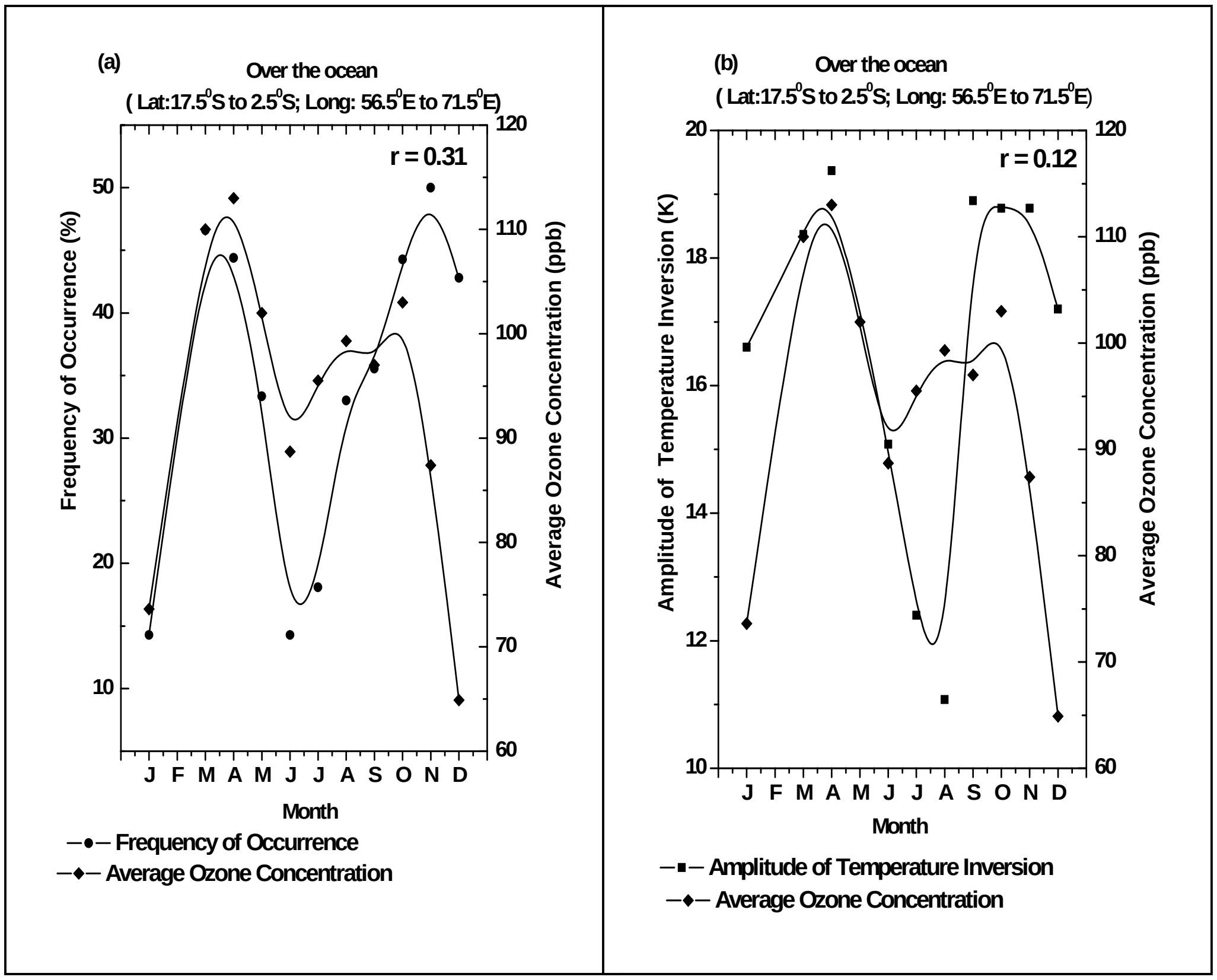
(a)
Over the ocean
( Lat:17.5$^0$S to 2.5$^0$S; Long: 56.5$^0$E to 71.5$^0$E)
r = 0.31
Frequency of Occurrence (%)
Average Ozone Concentration (ppb)
J F M A M J J A S O N D
Month
Frequency of Occurrence
Average Ozone Concentration
(b)
Over the ocean
( Lat:17.5$^0$S to 2.5$^0$S; Long: 56.5$^0$E to 71.5$^0$E)
r = 0.12
Amplitude of Temperature Inversion (K)
Average Ozone Concentration (ppb)
J F M A M J J A S O N D
Month
Amplitude of Temperature Inversion
Average Ozone Concentration

Figure 10